\documentclass[final,5p,times,twocolumn]{elsarticle}

\usepackage[utf8]{inputenc}
\usepackage{amssymb}
\usepackage{amsmath}
\usepackage{mhchem}
\usepackage{stfloats}
\usepackage{hyperref}
\usepackage{siunitx}
\usepackage{todonotes}
\usepackage{textgreek}
\usepackage{float}
\usepackage{wasysym}

\date{\today}
\journal{Fusion Engineering and Design (SOFT 2020, P1.22)}
\usepackage{color}
\usepackage{balance}
\begin{document}

\begin{frontmatter}

\title{Serpent neutronics model of Wendelstein 7‑X for 14.1 MeV neutrons}

\author[HGW]{Simppa \"Ak\"aslompolo$^\mathrm{a,}$\corref{cor1}}
\author[HGW]{Jan Paul Koschinsky}
\author[AALTO]{Joona Kontula}
\author[HGW]{Christoph Biedermann}
\author[HGW]{Sergey Bozhenkov}
\author[AALTO]{Taina Kurki-Suonio}
\author[VTT]{Jaakko Leppänen}
\author[AALTO]{Antti Snicker}
\author[HGW]{Robert Wolf}
\author[LLNL]{Glen Wurden}
\author{W7-X Team}
\cortext[cor1]{simppa.akaslompolo@alumni.aalto.fi}

\address[AALTO]{Department of Applied Physics, Aalto University, FI-00076 AALTO, FINLAND.}
\address[HGW]{Max-Planck-Institut f\"ur Plasmaphysik,
Teilinstitut Greifswald,
Wendelsteinstraße 1,
D-17491 Greifswald}
\address[VTT]{VTT Technical Research Centre of Finland Ltd, P.O. Box 1000, FI-02044 VTT, Finland}
\address[LLNL]{Los Alamos National Laboratory, P.O. Box 1663, Los Alamos, NM 87545, USA}

\begin{abstract}
In this work, a Serpent 2 neutronics model of the Wendelstein 7-X (W7-X) stellarator is prepared, and an response function for the Scintillating-Fibre neutron detector (SciFi) is calculated using the model.
The neutronics model includes the simplified geometry for the key components of the stellarator itself as well as the torus hall.
The objective of the model is to assess the \SI{14.1}{MeV} neutron flux from deuteron-triton fusions in W7-X, where the neutrons are modelled only until they have slowed down to \SI{1}{MeV} energy.
The key messages of this article are: demonstration of unstructured mesh geometry usage for stellarators, W7-X in particular; technical documentation of the model and first insights in fast neutron behaviour in W7-X, especially related to the SciFi: 
the model indicates that the superconducting coils are the strongest scatterers and block neutrons from large parts of the plasma.
The back-scattering from e.g. massive steel support structures is found to be small.
The SciFi will detect neutrons from an extended plasma volume in contrast to having an effective line-of-sight.

\end{abstract}


\end{frontmatter}

\section{Introduction}
In preparation for a future deuterium operation of the Wendelstein 7-X (W7-X) stellarator, potential neutron diagnostics must be evaluated.
The Scintillating-Fibre (SciFi) neutron detector~\cite{Wurden1995} can be used to detect the \SI{14.1}{MeV} neutrons originating from tritium burn-up process, where tritons from DD fusion fuse with the bulk deuterium:
\begin{align*}
\ce{^2_1D + ^2_1D ->  & ^3_2He + ^1_0n (\SI{2.45}{MeV})} & \si{50}{\%}\\
\ce{^2_1D + ^2_1D ->  & \underset{\ce{v}}{^3_1T} + ^1_1p & \si{50}{\%} }\\
                    \ce{^2_1D + &^3_1T -> ^4_2He + ^1_0n (\SI{14.1}{MeV})}.
\end{align*}
The large background of DD (\SI{2.45}{MeV}) neutrons and gamma rays form a challenging background noise for the measurement of DT neutrons.
By measuring the tritium burn-up, the fast ion confinement of W7-X can be assessed.
The fast ion confinement at high plasma pressure is one of the optimization targets of the W7-X design, and thus demonstrating it is a high-level project goal.

The state of the art neutronics model for W7-X is made by Gr\"{u}nauer~\cite{grunauer2008,grunauer2017} with MCNP.
While the materials are well defined, the geometry is highly simplified. 
The current work demonstrates the use of detailed geometry in the model,  which is achieved using the \emph{unstructured mesh}~\cite{serpent2cad4iter} based geometry type of Serpent 2~\cite{Leppnen2015}. 
The neutron scattering in material reduces the energy and randomizes the travel direction of the neutrons.
Thus, the scattering can disperse neutron flux traveling towards the detector (shielding) or reroute neutrons towards the detector (back-scattering), and thus detailed geometry in the proximity of the detector is important.
DT (\SI{14.1}{MeV}) neutrons reaching SciFi below approximately \SI{2.45}{MeV} energy are indistinguishable from DD-neutrons and are thus discarded at \si{1}{MeV} energy.

The SciFi detector consists of an array of scintillating plastic fibres inserted inside parallel holes drilled into an aluminium cylinder.
The high-energy neutrons scatter from the hydrogen in the plastic and the recoil protons induce scintillation in the fibre, which makes the detector insensitive to thermal neutrons.
The light is guided to a photo multiplier tube for amplification and later detection.
The parallel fibres induce a natural directionality to SciFi.
The detection efficiency as a function of neutron streaming direction was measured at the Physikalisch-Technische Bundesanstalt (PTB), Germany's national metrology institute, and is used as one of the inputs for this work.

The following section~\ref{sec:model} describes the Serpent~2 model created in this work, as well as how an response function for SciFi is calculated.
The tabulated function facilitates calculation of SciFi signals for various fusion rates without re-running the whole neutron transport calculation.
The section~\ref{sec:instrument} shows key features of the calculated response function and a first estimate of the SciFi signal from a hypothetical deuterium plasma is given in the section~\ref{sec:results}.
The paper is concluded by a discussion and summary in section~\ref{sec:discSum}.

%
%

\section{The W7-X model implementation in Serpent}
\label{sec:model}

A simulation model in Serpent consists of a description of materials, geometries, detectors and neutron sources.
Serpent uses the material isotopic compositions and densities while evaluating of scattering etc. cross sections. 
In this work, the Joint Evaluated Fission and Fusion (JEFF) Nuclear Data Library 3.1.1, as a copy is provided with Serpent~2 (pre-release version~2.1.32).
In the current model, the detectors are set up as track-length-detectors that tally the neutron flux within the detector, which is defined as the aluminum cylinder housing the scintillating fibres.  
The next section describes the geometry of the material boundaries and detector shapes used in the calculations.

\subsection{Geometry}
\begin{figure}[b]
  \centering
    \begin{minipage}[t]{0.49\linewidth}%
\includegraphics[width=\linewidth,trim=0mm 0mm 0mm 0mm,clip=true]{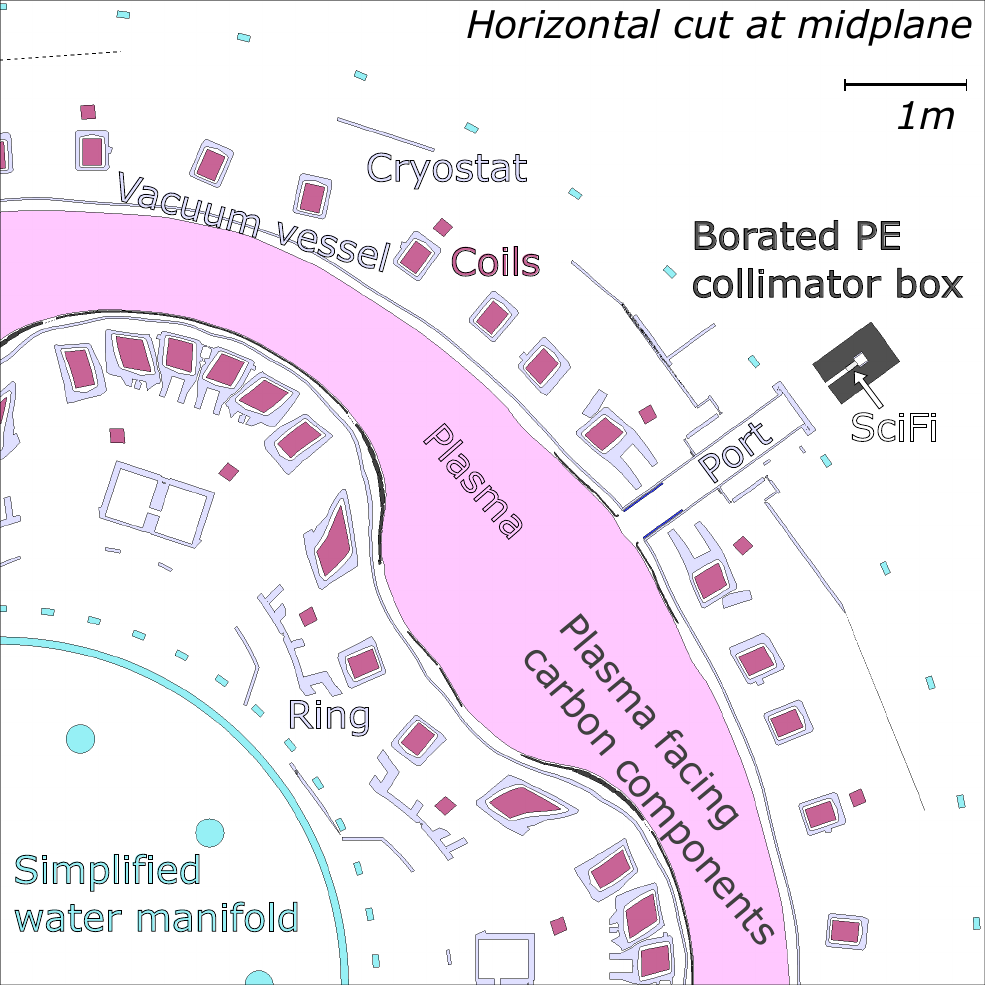}%
  \caption{The modelled geometry. The model is toroidally 5-fold symmetric. }
  \label{fig:geom}
\end{minipage}\hfill
    \begin{minipage}[t]{0.49\linewidth}%
\includegraphics[width=\linewidth,trim=0mm 0mm 0mm 0mm,clip=true]{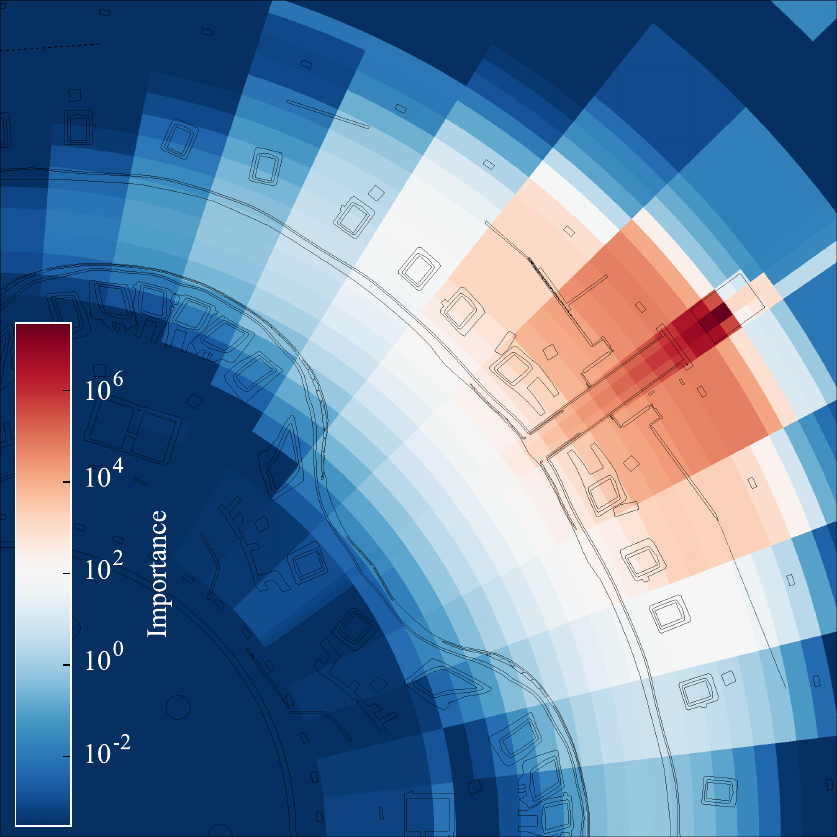}%
\caption{Variance reduction neutron importances used to concentrate the simulation effort to the neutron histories likely to contribute to SciFi signal.  }
  \label{fig:weightWindows}
 \end{minipage}
\end{figure}

The Wendelstein 7-X stellarator consists of hundreds of thousands of components and many of them are custom 3D-designs.
The computer aided design (CAD) model of the machine consists of \num{70000} unique parts, many of which have \num{10} instances.
The in-vessel parts cover \SI{260}{m^2} of surface, consists of about \SI{250000}{parts}, \num{4000} of which are major. All the surfaces are (or will be) actively cooled, with several kilometers of water lines in-vessel.
It is practically impossible to introduce the whole complex CAD model into any current neutronics tool and expect to have a sufficiently fast model.
Thus, it was necessary to pick the most significant components  and only include them.
Even these geometries required simplifications due to the extremely detailed 3D-designs of most components.

The model mainly consists of triangular meshes~\cite{serpent2cad4iter} representing CAD models exported from the CATIA design database as {\tt .stp}-files with minor details manually removed in CATIA.
The {\tt .stp}-files were then studied in FreeCAD, meshed using algorithms and parameters adapted for each component, and finally saved in {\tt .stl}-format.
Serpent includes the {\tt -checkstl} command line option for checking an {\tt .stl} for leaking or otherwise malformed surface meshes.
The leak-locations were imported back to FreeCAD, and the broken details healed or removed.
This was iterated until the meshed model was leak proof.

Further manual labor was needed for certain components. 
In particular, the inner vacuum vessel (VV) CAD model was unusable and, therefore, the VV model was recreated from simplified parametric representation lacking the port holes and the one for port AEU20 was recreated.
Water cooling of plasma facing components was included by making rough estimates on water/steel or water/carbon etc. ratios in the materials and creating homogenized materials. 
The very simple model for the water cooling manifold was adapted from~\cite{grunauer2008}.
The model geometry is 5-fold toroidally symmetric around the AEU20 port.

The model includes the following components (figure~\ref{fig:geom}): the plasma domain; plasma facing carbon components (PFC): divertors and shields; the inner vacuum vessel (VV); the non-planar coils with steel casing; the planar coils; the central support ring holding the coils; the AEU20 port and the port liner; the outer vacuum vessel (cryostat); the borated polyethylene (PE) collimator box around Scifi; an aluminium cylinder representing SciFi, and shielding components present inside the PE box. The whole model is inside a simple concrete torus hall.

There are of course many components not (yet) included, such as: water cooled steel panels; large nearby components in the torus hall, such as the neutral beam injection box; cables and pipes in the cryostat; and planar coil casings.

\subsection{Model optimization}
The Serpent geometry engine is very powerful, when used correctly.
The geometry can be defined as a combination of constructive solid geometry (CSG) and meshed surfaces from {\tt .stl}-files. 
In the current model, there are approximately~\SI{3}{million} triangles.
A good geometry definition divides the model into simple disjoint volumes to allow Serpent to quickly narrow down to the relevant components.
This was achieved by including the components in concentric \emph{universes} so that the most often used component (plasma domain) was always checked first.
Furthermore, Serpent uses search trees to expedite the mesh searching.
The optimal trees are sufficiently deep and branched to enable Serpent to skip the bulk of mesh checks and instead directly look up the active geometry part from the search tree.
Optimizing the depth and spatial divisions enabled a further speedup.

By studying the solid angles, it can be estimated that less than one out of \num{100000} neutrons would reach the detector even when born in the most favorable locations of the plasma.
Generating sufficient statistics in the whole plasma domain with reasonable energy resolution would take excessively long.
Thus, it is necessary to bias the simulation to predominantly follow neutron histories that are likely to reach the detector and predominantly ignore the others~\cite{MCPrtTransportMethodsNeutronPhotonCalculations} and 
Serpent uses standard Monte Carlo methods to achieve this~\cite{ResponseMatrixMethodImportanceSolverVarianceReductionSchemeSerpent2}.
To gain sufficient number of counts in the detector, the Serpent global variance reduction (VR) was used in a model without the PE collimator box and anisotropic detection efficiency (see next section).
Next, the weight windows were optimized for gathering counts in the SciFi detector, still without the directionality.
The resulting neutron importances were used in the production runs and are presented in figure~\ref{fig:weightWindows}.
The concept of importance can be interpreted as the contribution that a single particle at given position, energy and direction of motion has to a specified response, such as a physical reaction rate.

In the end, it was possible to calculate 868 histories per CPU-second from the initial energy of \SI{14.1}{MeV} to the cut-off energy of \SI{1}{MeV} (which is well below the DD birth energy). 
The analog (without variance reduction) simulation performance was 615 histories per CPU-second.
The code required a maximum of \SI{15}{GB} of memory per computing node.

\section{SciFi response function calculation}
\label{sec:instrument}
The main goal of this work is to implement a method for calculation of the SciFi response function or weight function.
That means a function that transforms a given fusion rate in a given location into the measured signal in SciFi.
To achieve this, \SI{14.1}{MeV}-neutrons are initialized uniformly with isotropic velocity distribution in the plasma domain.
Their histories are followed in the above described geometry using the variance reduction methods.

In the SciFi, a track-length-detector tallies the average neutron flux inside the SciFi:
\begin{equation}
    \Phi_\mathrm{det}=\frac{1}{V_\mathrm{det}}\int_\mathrm{tracks}\theta w\mathrm{d}l,
\end{equation}
where $w$ is the neutron marker weight depicting how many real neutrons per second the marker represents.
The anisotropic detection efficiency, $\theta$, of SciFi scales the neutron flux as a function of the angle between the SciFi axis and the neutron track\footnote{A Serpent 2 feature implemented by the author for this work.}.
The whole SciFi aluminium cylinder is used as a proxy for the active media, the fibres, to increase the integration volume.
The flux $\Phi_\mathrm{det}$ is binned in a cylindrical grid as a function of the marker birth location\footnote{Another Serpent 2 feature implemented by the author for this work.}. 

From the flux one gets to the counts by integrating an (arbitrary) neutron production rate $S$  weighted by the response function over the plasma volume.
\begin{equation}
    \frac{\mathrm{counts}}{\mathrm{s}}=\sum_\mathrm{bins}\int_\mathrm{bin}\underbrace{A_\epsilon\Phi_\mathrm{det}\frac{ V_\mathrm{plasma}}{NV_\mathrm{bin}}}_\mathrm{response\ function}S\mathrm{d}V,
    \label{eq:counts}
\end{equation}
where $A_\epsilon$ is the effective active area of the SciFi detector measured experimentally at PTB~\cite{PTBmeasurements} (\SI{1.2}{mm^2}), and is simply the efficiency of which SciFi countrate results from a certain neutron flux into the detector. 
Ideally,  $A_\epsilon$ and $\theta$ would be measured as a function of energy, but in this work only the value for \si{14.8}{MeV} is used.
Serpent launched $N$ neutrons per second in the whole plasma volume $V_\mathrm{plasma}$, which needs to be distributed to the bins according to their volumes ($V_\mathrm{bin}$).

\begin{figure}[H]
  \centering%
\includegraphics[width=0.499\linewidth,trim=58mm 195mm 355mm 43mm,clip=true]{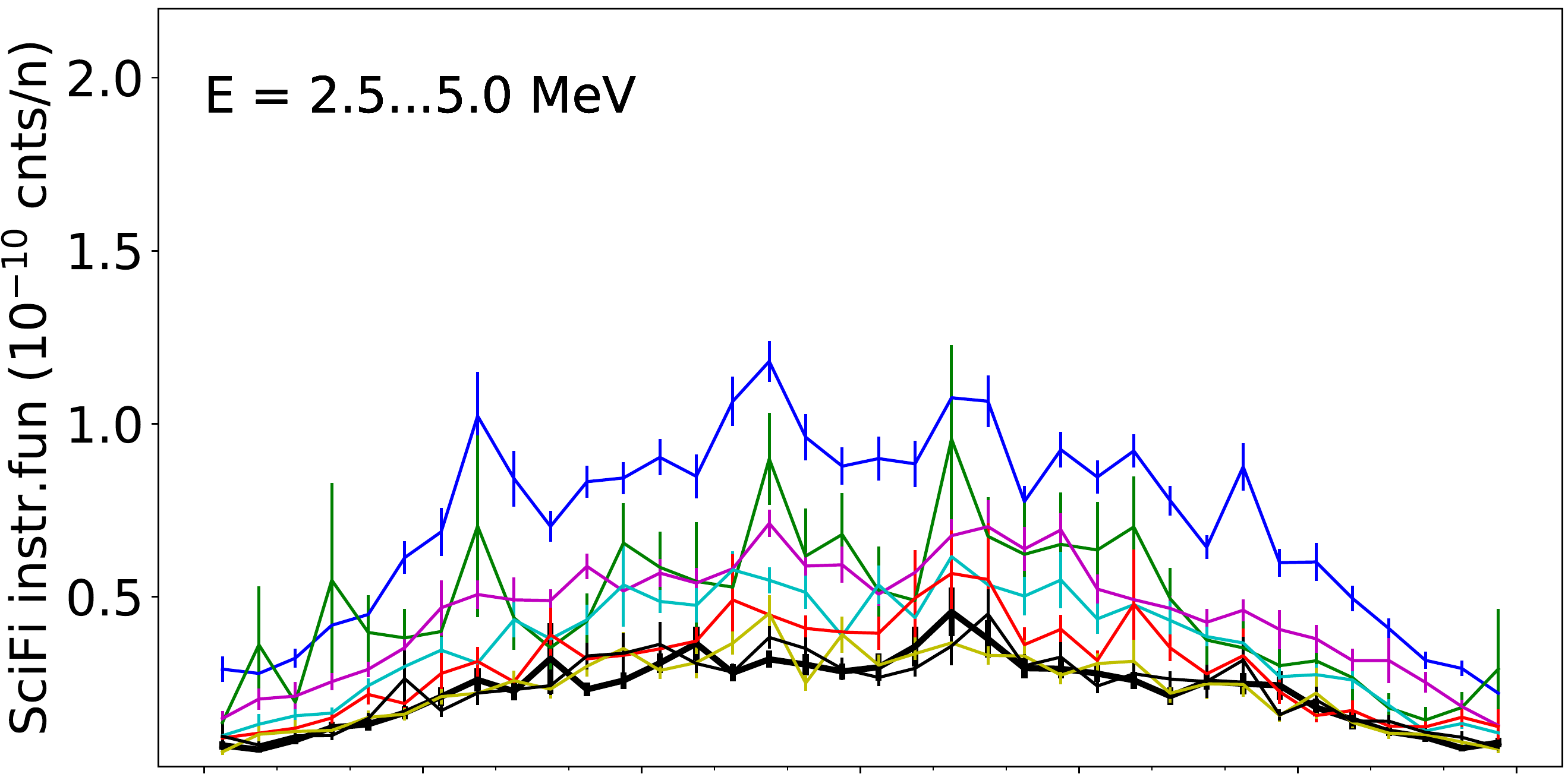}%
\includegraphics[width=0.499\linewidth,trim=372mm 195mm 41mm 43mm,clip=true]{fusionRates_on_axis_poster6.pdf}\\
\includegraphics[width=0.499\linewidth,trim=58mm 19mm 355mm 199mm,clip=true]{fusionRates_on_axis_poster6.pdf}%
\includegraphics[width=0.499\linewidth,trim=372mm 19mm 41mm 199mm,clip=true]{fusionRates_on_axis_poster6.pdf}%
  \caption{The detection efficiency ($\pm1\sigma$) near the magnetic axis as a function of toroidal angle for various energy ranges. Similarly with various components omitted in the model. The detector is at \SI{36}{\degree}.
  }
  \label{fig:instrFunAxis}
\end{figure}

\section{Simulation results}
\label{sec:results}

The response function was calculated by launching \num{4.5e9}~neutron histories.
A one or two orders of magnitude larger simulation would have been technically possible using the Cobra supercomputer at Max-Planck computing \& data facility, but the current Monte Carlo noise and spatial resolution was deemed sufficient for the current study.
The error bars shown are the statistical errors in the Monte Carlo simulation, and do not include errors in inputs or other systematic errors, most notably the missing components, but also i.a. possible issues in material definitions and nuclear data.
To assess the relative significance of the various components, the calculation was repeated with certain components removed.

The resulting response function is stored in a regular cylindrical grid with vertical and radial grid pitch of~\SI{10}{cm} and toroidal pitch of~\SI{2}{\degree}. 
The fusion rate is expected to peak near the stellarator magnetic axis, so a natural visualization is to show the response function along the axis as a function of the toroidal angle. Figure~\ref{fig:instrFunAxis} shows this for four different neutron track energies and for the calculations with specific components removed.
The detection efficiency is toroidally wide, i.e. the "line-of-sight" through the port and collimator is not evident.
This is explained by noting that approximately half of the counts are due to neutrons with energy below \SI{10}{MeV}, i.e. after scattering in matter before reaching the detector. Also the PE collimator dimensions indicate a wide cone: a \SI{20}{cm} deep, \diameter \SI{4}{cm} collimator will lead to an approximately \SI{50}{cm} diameter acceptance cone at \SI{3}{m} distance.
The various components all act to reduce the neutron flux to the detector.
For scattered neutrons, the coils are the largest shadowing element.
Back-scattering from the ring or water containing plasma facing components (PFC) appears small.

\begin{figure}[t]
  \centering
  \includegraphics[width=1.0\linewidth,trim=17mm 13mm 45mm 30mm,clip=true]{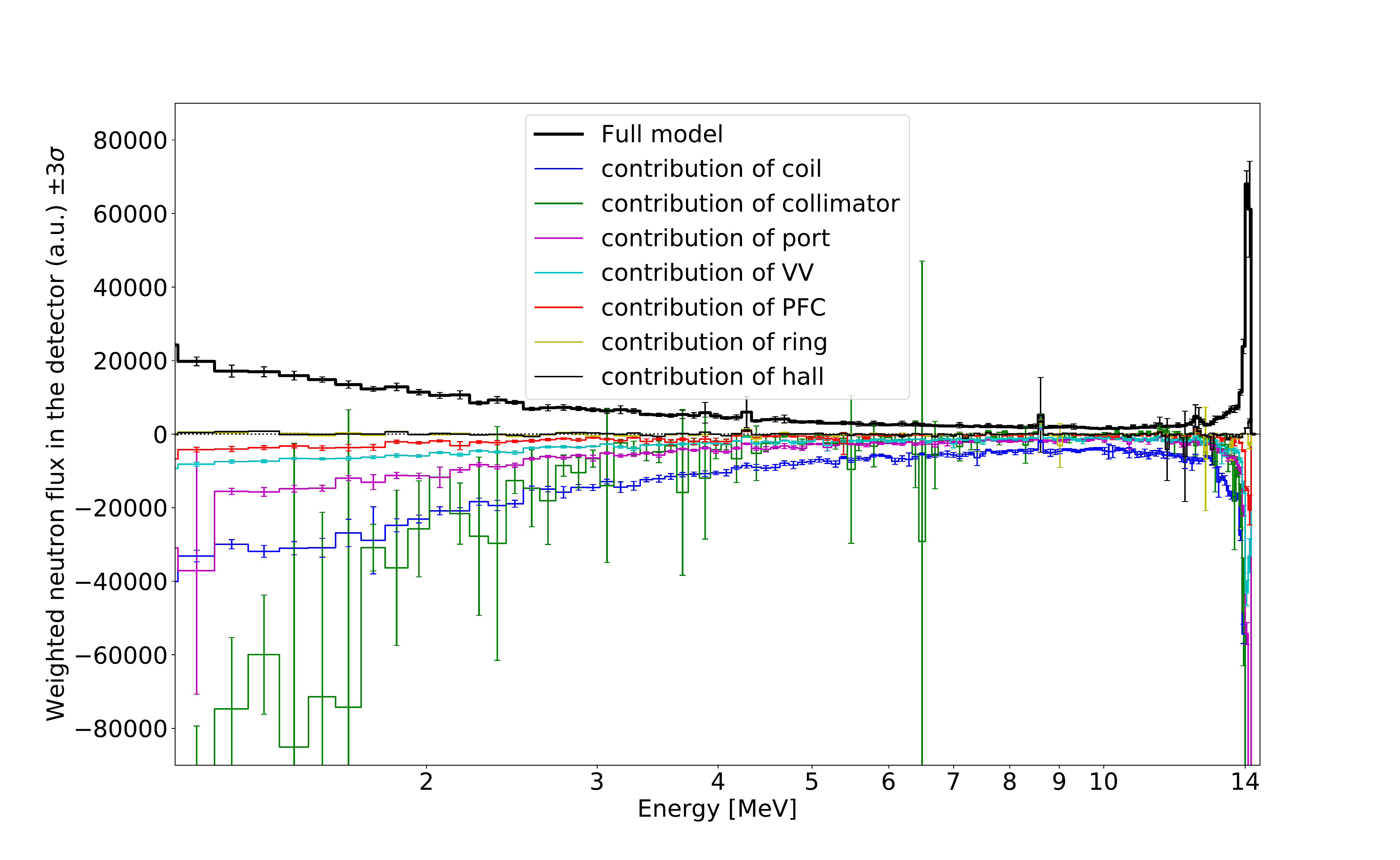}
  \caption{The neutron flux in the detector weighted with the anisotropic detection efficiency as a function of neutron energy.
  The neutrons are all born with \SI{14.1}{MeV} energy and most reach the detector only after scattering to lower energy.
  The colored lines depict the difference to the "Full model" when removing the indicated components from the simulation: $\Phi_\mathrm{contribution} = \Phi_\mathrm{full\ model}-\Phi_\mathrm{partial\ model}$.
  For most energies, the coils are the strongest to shield neutrons from the SciFi, except for lowest energies where the borated polyethylene collimator has the largest effect.
  } 
  \label{fig:Espectrum}
\end{figure}

If the spatial dependency is dropped, it is possible to significantly increase the energy resolution and study the energy spectrum of neutrons reaching the SciFi detector, as illustrated in figure~\ref{fig:Espectrum}.
It is obvious that only a tiny fraction of neutrons reach the detector without scattering while in transit.
No component increases the flux significantly, not even the massive steel ring on the opposite side of the plasma to the detector.
A minor back-scattering signal (positive values in  figure~\ref{fig:Espectrum}) may happen due to scattering in the concrete torus hall walls or in the collimator.

A ten-fold larger analog simulation was performed, to check for biasing caused by the VR.
The VR energy spectrum was found to be consistent with analogous calculation but the detector efficiency near axis had insufficient statistics to draw any conclusions. 

\begin{table}[b]
    \centering
\begin{tabular}{lllllllllll}
                 &            &            &            &           &            &           &            &           &            &               \\
\emph{Bin limits (MeV)} & \multicolumn{2}{c}{2.5} & \multicolumn{2}{c}{5.0} & \multicolumn{2}{c}{10.0} & \multicolumn{2}{c}{12.5} & \multicolumn{2}{c}{14.4}     \\\hline
\emph{Counts/s}         &            & \multicolumn{2}{|c|}{40}  & \multicolumn{2}{c}{35}   & \multicolumn{2}{|c|}{16}   & \multicolumn{2}{c|}{67}   &            
\end{tabular}
    \caption{The energy spectrum of the neutron counts (158 tot). The results show number of DT neutrons per second detected by SciFi. The energy depicts the energy of the originally \SI{14.1}{MeV}-neutrons when it arrives to the detector. (The \SI{14.8}{MeV} detection efficiency measured at PTB is used for all energies~\cite{PTBmeasurements}.)}
    \label{tab:fullStackSpectrum}
\end{table}
The ultimate use of the response function is to estimate SciFi signal due to a real plasma using equation~\ref{eq:counts}.
As a proof of concept, the fusion rate is calculated using the ASCOT suite of codes, as described in~\cite{ASCOTW7XneutronTatesMagneticConfiguration}: The neutral beam injection of deuterium is modelled with BBNBI, their slowing down with ASCOT5, the ensuing D(D,p)T reaction rate is calculated with AFSI. The triton slowing down is again calculated with ASCOT5 and AFSI then calculates the D(T,\textalpha)n reaction rate.
Serpent then calculates the neutron flux at the SciFi detector. 
The flux is converted to counts using the anisotropic detection efficiency calibration measured at PTB~\cite{PTBmeasurements}.

The result is \SI{158}{counts/s} from triton burnup neutrons in a hypothetical Wendelstein 7-X plasma heated by 8 NBI sources (\SI{12}{MW} tot) having axial density of  $n$=\SI{2e20}{1/m^3} and temperature of $T$=\SI{1.4}{keV}  producing \SI{1.6e12}{n/s}.
        Thus, an estimated SciFi integration time could be 50 ms.
        Table~\ref{tab:fullStackSpectrum} shows how these counts are distributed to different energy ranges.
 The \SI{14.8}{MeV} detection efficiency is used for all energies. A more refined energy dependent calibration would likely lead to 
higher count rates. 
With the current settings for pulse discrimination, the effective active area of SciFi increases when the neutron energy drops from \SI{14.8}{\mega eV} to \SI{2.49}{MeV} and the directionality is reduced.

\section{Discussion and Summary}
\label{sec:discSum}

This paper demonstrates that it is possible to generate a Monte Carlo neutronics model of a complex 3D shaped stellarator using the Serpent 2 code.
The first author was able to create the model in approximately three months without any prior knowledge of Serpent. 
Each component took typically two working days (with large variance) of manual labor to prepare for Serpent usage. 
The model is used to calculate an response function for a scintillating fibre neutron detector.
The detector is found to be wide-angle, in the sense that it detects \SI{14.1}{MeV} neutrons from a large volume, not only from a "line-of-sight".
This is because most neutrons reach the detector only after scattering. 
However, little back-scattering of neutrons is found, meaning, removal of any component from the model increases the counts in SciFi.
This would imply, that by adding further components the model will further reduce the SciFi signal, and thus the integration time estimate for SciFi (\SI{50}{ms}) is probably a lower bound.
All in all, this analysis is a preliminary study, and this article is primarily technical documentation of the methods. 

Future work includes introducing missing components and applying the model to various calculated neutron rates to assess the capability of the system to detect variations in the neutron rate caused by various experiments.
Further validation would also be beneficial.

\section*{Acknowledgements}
\noindent This work has been carried out within the framework of the EUROfusion Consortium and has received funding from the Euratom research and training programme 2014-2018 and 2019-2020 under grant agreement No 633053. The views and opinions expressed herein do not necessarily reflect those of the European Commission. This work was partially funded by the Academy of Finland project No. 328874 and No. 324759.  The authors wish to acknowledge CSC – IT Center for Science, Finland, for computational resources as well as the Max-Planck computing \& data facility for the same. The HPCE$^3$ programme and Walter Ahlström foundation contributed to the funding of the work. The authors want to thank the operators and team at PTB for support during the measurements performed at the PTB ion accelerator facility, especially E. Pirovano, A. Lücke, M. Zboril and A. Zimbal.

\bibliographystyle{elsarticle-num} 
\bibliography{bibliography}

\end{document}